# A COMMENT ON THE LLA METHOD, THE $k_T$ JET ALGORITHM AND THE BFKL THEORY

Talk at the Joint international workshop "Hadron Structure and QCD" (HSQCD'2008), Gatchina, 30 June — 4 July, 2008.

F. V. TKACHOV
Institute for Nuclear Research of Russian Academy of Science,
Moscow, 117312, Russian Federation
E-mail: ftkachov@ms2.inr.ac.ru

## Abstract

The leading logarithmic approximation method fails to yield the correct asymptotic behavior in some realistic situations: inclusion of the $\beta$-terms to which the LLA method is insensitive may change a power growth to merely logarithmic. The results of [hep-ph/0101058] indicate that the problem of large-$s$ behavior of total cross sections belongs to this class. Similarly, a reference to the LLA method cannot be sufficient to justify constructions such as the $k_T$ jet algorithm.

## Introduction

This talk is essentially a postscriptum to [1]. At the symposium [2], I reported results of my investigation of some pQCD problems with the method of Asymptotic Operation (AO; for references and comments see below). For total cross sections, I quoted a factorization theorem that had a few asymptotically dominant logarithmic functions satisfying a purely differential evolution equation of a conventional form with the standard $\beta$-terms. However, while still at the workshop, some terms where recovered that were lost on the way from intermediate to final formulae (because the structure of factorization is somewhat unusual in this case), and that spoiled the simplicity of the final formulae [1]. However, the disappointment at the recovery of the spoiler terms prevented me from appreciating another detail of the derivation, and from realising that, although the May, 2000 conclusions were overoptimistic, they were not totally wrong.

The conclusion should be that in a complete (beyond-LLA) treatment, the standard $\beta$-terms emerge in the asymptotic evolution equations in a natural way, and irrespective of any additional complications, and such terms are what is called a ***singular perturbation***. This means that, although such terms are technically of a higher order in the coupling (i.e. correspond to non-leading logarithms), they induce a qualitative change in the asymptotic behavior. In the case of total cross sections this implies that the power behavior is converted into a logarithmic one, the latter unobtainable from the former by a small correction (i.e. the limit $\beta \to 0$ is non-analytic). It is then impossible to avoid the conclusion that this apparently invalidates the entire mythology around the LLA method in general and the power behavior of total cross sections (reggeization etc.) in particular. It should also be noted that attempts to ascribe physical significance to properties such as magnitude of the terms are an exercise in futility if the series is divergent as is the case with resummations of large logarithms in perturbative quantum field theory.

## $\beta$-terms as a singular perturbation

Take a quantity, say, $X(\alpha,\mu)$, $\alpha$ and $\mu$ being the coupling and the renormalization/factorization parameter, respectively. Assume that it is represented by a sum of perturbative integrals (a power series in $\alpha$). Suppose that $X$ satisfies a standard evolution equation:

$$\boxed{\left(\partial_L + \beta\partial_\alpha\right)} X = \gamma X, \quad L = \ln\mu^2 \qquad (1)$$

where the frame encloses the standard differential evolution operator, and where

$$\beta(\alpha) = -\beta_0\alpha^2 + O\left(\alpha^3\right), \quad \beta_0 > 0; \qquad \gamma(\alpha) = \gamma_0\alpha + O\left(\alpha^2\right)$$

For instance, moments of DIS structure functions satisfy such equations.

The asymptotic solution at large $\mu$ (or small $\alpha$, due to the positivity of $\beta_0$) is easily obtained and is well-known:

$$X(L) \sim \left(\ln\mu^2\right)^{-\left(\gamma_0/\beta_0\right)} \qquad (2)$$

One sees that:

1) The asymptotic behavior is logarithmic.

2) It is independent of $\alpha$.

3) The limit $\beta \to 0$ is singular.

Now, the **LLA method** (which stems from Sudakov's seminal work [3]) is notoriously insensitive to the $\beta$-terms (cf. the usual difficulties with incorporating the running coupling into the results), and is essentially equivalent in this case to solving

$$\boxed{\left(\partial_L\right)} X = \gamma X, \quad L = \ln\mu^2 \qquad (3)$$

The difference from the exact equation is in the simplified form of the differential operator in the frame on the l.h.s.

The asymptotic solution is immediately seen to be as follows:

$$X(L) \sim \left(\mu^2\right)^\gamma \qquad (4)$$

That the $\beta$-term in the exact differential operator is a *singular perturbation* is easily seen: the two asymptotic expressions for $X(L)$ cannot be formally related by setting $\beta = 0$.

Such power behavior is characteristic of the LLA treatment of the total cross sections [4], [5] and has given rise to a vast field of theoretical activity. However, within the LLA treatments, the power behavior is established by direct resummation of leading logarithms, perhaps organized into an integral equation, rather than solving differential evolution equations. The differential evolution equations with a $\beta$-term are a feature that is foreign to the LLA treatments (where instead things revolve around integral equations such as the BFKL one [5]), and the transition from the power behavior (4) to the logarithmic one (2) is by no means apparent. However, the analysis of [1] establishes a connection with such evolution equations, and the above argument is thus actualized:

— A complete (systematically all-log) treatment of the total cross sections at large $s$ [1] shows presence of the $\beta$-terms in the evolution equations. Irrespective of whether or not those equations can be written in a convenient closed form, the $\beta$-terms are there.

— Therefore, the power behavior of total cross sections is an artefact of LLA.

— The logic dictates that in whatever speculations about total cross sections' asymptotic behavior one wishes to be engaged, the asymptotic behavior should by default be taken logarithmic where the power behavior $s^{c\alpha}$ is customarily used.

I don't see how to avoid these conclusions, or whether one should try to.

## A note on jet algorithms

If Sudakov's LLA method may/does yield wrong asymptotic behavior, then its special physical significance is only a myth, and any conjurations to it in specific problems lack substance unless further evidence is produced. (In fact, one ought to regard too tight a connection with the LLA method a drawback rather than an advantage.)

In particular, one concludes that the argumentation behind the $k_T$ algorithm [6] is insufficient. Which leaves us with the Optimal Jet Definition [7] as the only jet definition available that was developed in a systematic manner on a scientific foundation.

(**Note.** Since there are misinterpretations in the literature, it is appropriate to emphasize that the OJD is defined independently of any implementation. Ref. [8] considered two implementations of OJD, and one must indicate the one used in any speed comparison, unlike was done in [9]. Anyhow, the speed of an implementation cannot be an argument per se, without regard to physical correctness. Lastly, the OJD is an integral part of a significant theory [7] that promulgates a systematic use of weights as an alternative to the standard methods such as maximal likelihood in HEP situations where it is typical to have a very large dimensionality of event spaces, and the probabilities are specified by MC event generators rather than explicit pdf's. Incidentally, the idea that weighting techniques ought to be given more prominence in HEP data processing [7] in the frequent situations where signal and background are not clearly separated, was seconded by a leading expert in jet algorithms and statistical methods [10].)

## Notes on the factorization theorem of [1]

Let $\Sigma(s)$ be a total cross section normalized by a power of $s$ to be dimensionless. One has to study its behavior at large $s$. A naive factorization (=parton model) result disregarding soft singularities would be

$$\Sigma(s) = \int_0^1 \frac{dz_1}{z_1} K_\mu(z_1) \int_0^1 \frac{dz_2}{z_2} K_\mu(z_2) \times \tilde{\Sigma}_\mu(z_1 z_2 s) + O\left(s^{-1}\right) \tag{5}$$

where $K_\mu(z)$ are parton distributions and $\tilde{\Sigma}_\mu(s)$ is the hard parton cross section. However, the integrals near $z=0$ are divergent, and a correct factorization result would be

$$\Sigma_{as}(s) = \Delta_\mu(s) + \int_c^1 \frac{dz_1}{z_1} K_\mu(z_1) \int_c^1 \frac{dz_2}{z_2} K_\mu(z_2) \times \tilde{\Sigma}_\mu(z_1 z_2 s) \tag{6}$$

where $c$ is a cutoff, and $\Delta$ depends on the parton distributions. The latter dependence was missed in my talk at [2] and corrected in [1]. However, ref. [1] still missed that it is actually possible to split $\Delta$ as follows:

$$\Delta_\mu(s) = \bar{\Delta}_\mu(s) + \bar{\bar{\Delta}}_\mu(s) \tag{7}$$

This split plays a key role in the following discussions, and it is important to understand its meaning well.

The first term on the r.h.s. of (7), $\bar{\Delta}_\mu(s)$, contains all the leading logarithms as well as some non-leading ones. It depends on $c$ but not on the shape of parton distributions (in the diagram-by-diagram analysis, it depends on neither partons' virtuality nor their masses). Diagrammatically, it is directly seen to correspond to the BFKL ladder (more precisely, it contains — along with a full set of appropriate radiative corrections — an asymptotic form of the BFKL ladder with the singularities properly subtracted according to the distribution-theoretic rules of the theory of AO).

The second term, $\bar{\bar{\Delta}}_\mu(s)$, contains only non-leading logarithms and depends on the shape of parton distributions, and therefore is treated as a non-perturbative entity similar to the matrix elements of local operators in the pQCD treatment of moments of DIS structure functions.

One could argue that due to the non-perturbative nature of $\bar{\bar{\Delta}}_\mu(s)$ and unlike the usual perturbative expansion (where taking into account subleading terms resolves ambiguities of scale etc.) — it is then a matter of taste whether to select and resum only leading logarithms (thus obtaining the power behavior) or to deal with $\bar{\Delta}_\mu(s)$ that includes some non-leading logarithms and satisfies an evolution equation of a conventional type and thus behaves logarithmically at large *s*. However, the separation into leading and non-leading logarithms is ambiguous, whereas the split (7) emerges in a natural fashion within the method employed (i.e. the method of AO), without special adjustments or manipulations.

I will come back to this "naturalness" argument below, but for the purposes of the present postscriptum it is sufficient to note that even if the logarithmic behavior emerges as an option, the power behavior is no longer a given, and is in fact the least natural choice.

The above factorization theorem can be illustrated as follows:

$$\Sigma(s) = \boxed{O\left(\ln^\Gamma s\right)} + \boxed{\bar{\bar{\Delta}}_{\mu,c}(s)} + \boxed{\text{parton model formula with cutoff at small } z\text{'s}} \tag{8}$$

The leading contribution is shown to behave logarithmically but remember that it incorporates, besides all the leading logs, also some sub-leading logarithms. Actually, a matrix (and mixing) is involved here:

$$O\left(\ln^\Gamma s\right) = \sum_{i,j} g_i^{h_1} g_j^{h_2} C_{ij}(s) \tag{9}$$

where (only) $g_i^h$ depend on global parton content of the hadron, but not on the specific shape of the parton distributions. Correspondingly, the anomalous dimension $\gamma$ that figures in the evolution equation for $C_{ij}$ is a matrix, so eigenvalues play a role in the asymptotic evolution, etc. — similarly to the case of DIS moments of structure functions (with appropriate implications for e.g. the Pomeranchuk theorem etc.)

(**Note.** In the case of total cross sections, the evolution equation might also contain an "anomalous addendum", i.e. the r.h.s. becomes $\gamma X + \delta$; at least terms to that effect tend to emerge in analysis of individual perturbative integrals although they tend to cancel in the sum total. A non-zero $\delta$ would not affect the logarithmic behavior (2).)

As was pointed out in [1], one could play games with varying, *s*-dependent cutoff *c*. This might allow one to asymptotically suppress the bad non-perturbative terms at large *s*, but it is unclear how this could be achieved in practice; perhaps, a dose of phenomenology might help.

### Comments on Asymptotic Operation

I have made a statement that the split (7) is natural within the method of AO. But the separation of leading logs that leads to the power behavior is also natural — within the LLA method. The issue would have been settled if it were possible to factor out also all non-leading logs. Unfortunately, this is not the case because there are non-perturbative contributions at the level of non-leading logs, and it is not clear whether the separation of "perturbative" non-leading logs to be included into the resummation from the "non-perturbative" ones to be left in the form of a phenomenological function of *s*, can be done uniquely. So I must put some evidence behind the assertion that the AO split is more natural than the LLA split.

Quite simply, the method of AO is radically more powerful than the LLA method: the former was used (to take just one example) to develop formulae for OPE [11] that made possible a number of NNLO calculations in a straightforward, well-automated fashion (the first such calculations are [12]) — a feat impossible with the LLA method. Moreover, the method of AO was designed to systematically develop complete asymptotic expansions of perturbative integrals, with all log and power terms taken into account, not just the leading logs in the leading power, and in a form suitable for studies of factorization, i.e. with a minimal mangling of the structures of the integrals being expanded. All this means that the method of AO reflects the underlying physical structures better than the method of LLA.

Of course, the LLA method is much older than the method of AO (1956 vs. 1982), it was used in a number of important calculations, and it developed a large following, publicity and, of course, mythology. This means that it will be around for perhaps another twenty years or so. But this has nothing to do with the intrinsic values of the two methods (cf. [13]).

References to the original discovery publications for the Asymptotic Operaction with some history comments can be found in [14]. Below I attempt to convey an idea of where the power of AO comes from.

The most important feature to understand about the method of AO is that it is based on the concept of generalized functions (distributions), and that's where its analytical power comes from. Formal treatments of the theory of distributions are not prerequisite to understand it, in practice one needs to learn just a few basic principles and technical tricks that can be learned from papers the original papers (see [1] and [14] for references).

Recall the famous ordering of partons

$$x_1 < x_2 < \ldots < x_n \qquad (10)$$

This sector decomposition due to Sudakov [3] (rediscovered by K.Hepp in the 60s in the context of the BPHZ theory of UV renormalization) is an essential splitting of the integration region that allows one to classify and extract asymptotic contributions. In the method of AO, one (sort of) focuses on just one link of this chain, with the tail hidden within a distribution. To take analogy with computer programming, this is like comparing an algorithm containing a loop with and equivalent recursive algorithm. The role of the apparatus of distributions is exactly to allow an efficient use of the recursive pattern in the problem (graph $\to$ subgraph $\to \ldots$). Whereas the treatment of a single link of the chain becomes similar to (and in appropriate notations indistinguishable from) the handling of the one-loop case.

Another point that AO exploits to obtain smooth and efficient expansion results, is that it is not only (perhaps, even not so much) the *regions* that are responsible for logarithms etc. but (rather) the *integrands*. And a flexible handling of the integrands (whose some parts are interpreted at intermediate steps as test functions for the other parts) is another important element of the technique.

A complication in the problems involving essentially non-euclidean asymptotic regimes (to which class belongs the problem of asymptotic behavior of total cross sections) is that a secondary expansion is necessary [15], [16] (such expansions occur in the finding of explicit expressions for the coefficients of $\delta$-functional counterterms that have to be added to the formal Taylor expansion of the integrand in the small parameter). In the case of total cross sections, the secondary expansion (the second branch of recursion) is associated with the soft singularities. Such secondary expansions introduce a second branch of recursion, and non-trivial double-branching recursions are usually hard to represent in terms of loops (as any experienced programmer knows). It should immediately be clear why the LLA method encountered difficulties in this case: it is very much like trying to code a non-trivial recursion algorithm in Fortran IV — possible but extremely cumbersome, and is only possible when the recursive solution is already known.

(**Note.** The latter point fully explains the idea behind the so-called "method of regions": the idea is to take a solution found by a superior novel method and rewrite it in an obsolete but well-established formalism, and exploit the fact that the novel method is ill-known to the masses to spread desinformation about it.)

Another beautiful aspect of the distributions-based techniques of AO is that singular products involving $\delta$-functions are treated in the same fashion as the corresponding products with ordinary propagators:

$$\frac{1}{(k-p_+)^2}\times\frac{1}{(k-p_-)^2} \leftrightarrow \frac{1}{(k-p_+)^2}\times\delta\left((k-p_-)^2\right) \tag{11}$$

Within the conventional techniques it is necessary to integrate out such phase space $\delta$-functions prior to starting any analysis. However, eliminating them obfuscates the structure of the integrands, making it much harder to see the true structure of the final factorization results — including the evolution equations they obey, such as eq. (1).

This preservation of the structure of integrands within the method of AO is why I believe the factorizations it yields (such as the one discussed above in connection with the evolution equation (1)) is more natural than the one that the LLA method yields. Whence my conclusion that the true asymptotic behavior of (properly normalized) total cross sections within perturbative quantum field theory is logarithmic (eq. (2)) rather than power-like (eq. (4)).

*Acknowledgments*. I thank V.S. Fadin for an update and L.N. Lipatov for the invitation to attend this workshop. I also thank (chronologically) L.N. Lipatov, B.I. Ermolaev and T.T. Wu for insightful discussions at earlier times, and CERN theorists for their hospitality that made these discussions possible. This work was supported in part by the grant of the President of the Russian Federation NS-1616.2008.2.

**References**

[1] F.V. Tkachov, *On the large-s behavior of total cross sections*. arXiv:hep-ph/0101058.
[2] International symposium *Evolution Equations and Large Order Estimates in QCD*, Gatchina, 30 April - 4 May, 2000.
[3] V. Sudakov, Zh. Eksp. Teor. Fiz. 30, 87 (1956) [Sov. Phys. JETP 3, 65 (1956)].
[4] H. Chen, T.T. Wu, Phys. Rev. Lett. v. 24 (1970), p. 1456-1450.
[5] L.N. Lipatov, *Small-x physics in perturbative QCD,* arXiv:hep-ph/9610276.
[6] S. Catani et al., Phys. Lett. 269B (1991) 432.
[7] F.V.Tkachov, *A theory of jet definition*. arXiv:hep-ph/9901444.
[8] D. Yu. Grigoriev, E. Jankowski, F. V. Tkachov, arXiv:hep-ph/0301185.
[9] M. Cacciari, G.P. Salam, arXiv:hep-ph/0512210.
[10] R. Barlow, arXiv:physics/0311105.
[11] F.V. Tkachov, Phys. Lett. 124B (1983) 212; S.G. Gorishny, S.A. Larin and F.V. Tkachov: Phys. Lett. 124B (1983) 217.
[12] S.A. Larin, F.V. Tkachov and J.A.M. Vermaseren, Phys. Rev. Lett. 66 (1991) 862.
[13] T. Kuhn, *The Structure of Scientific Revolutions*, University of Chicago Press, 1962.
[14] F.V. Tkachov, *Distribution-theoretic methods in quantum field theory*, arXiv:hep-th/9911236.
[15] F.V. Tkachov, Phys. Atom. Nucl. 56 (1993) 1558-1572.
[16] F.V. Tkachov, *Towards systematic near-threshold calculations in perturbative QFT*, arXiv:hep-ph/9703424.